\documentclass[twocolumn,prd,nofootinbib,showpacs,floatfix,preprintnumbers]{revtex4}
\usepackage{epsfig}
\usepackage[dvips]{color}

\renewcommand{\d}{\textrm{d}}
\newcommand{\der}[2]{\frac{{\d}#1}{{\d}#2}}   
\newcommand{\mchi}[0]{m_\chi c^2}
\newcommand{\mchiq}[0]{\left(\frac{\mchi}{\textrm{GeV}}\right)}
\newcommand{\sigva}[0]{\left\langle \sigma_{\text{A}}v\right\rangle}
\newcommand{\nuq}[0]{\left(\frac{\nu}{\textrm{Hz}}\right)}
\newcommand{\nuz}[0]{\left(\frac{\nu_0}{\textrm{Hz}}\right)}

\newcommand{\rclo}{\left(\frac{r_{\text{cl}}}{\textrm{kpc}}\right)}
\newcommand{\dcl}{\left(\frac{d_{\text{cl}}}{\textrm{kpc}}\right)}

\newcommand{\B}{\left(\frac{B}{\mu\textrm{G}}\right)}

\newcommand{\Ee}[0]{\left(\frac{E_{\text{e}}}{\textrm{GeV}}\right)}

\newcommand{\rhocl}{\left(\frac{\rho_{\text{cl}}}{\textrm{GeV$c^{-2}$cm}^{-3}}\right)}

\newcommand{\ms}[1]{10^{#1}M_{\odot}}

\newcommand{\moMS}[1]{10^{#1} M_{\odot}}

\def\alt{\raise0.3ex\hbox{$\;<$\kern-0.75em\raise-1.1ex\hbox{$\sim\;$}}}
\def\agt{\raise0.3ex\hbox{$\;>$\kern-0.75em\raise-1.1ex\hbox{$\sim\;$}}}

\definecolor{Black}{named}{Black}
\definecolor{Red}{named}{Red}

\newcommand{\bw}{\begin{widetext}}
\newcommand{\ew}{\end{widetext}}
\def\d{{\rm d}}

\begin{document}
\preprint{DSF/21/2008, IFIC/08-48}
\title{Radio constraints on dark matter annihilation in the galactic halo and its substructures}
\author{
E.~Borriello$^{1,}$\footnote{eborriello@na.infn.it},
A.~Cuoco$^2$\footnote{cuoco@phys.au.dk}
and G.~Miele$^{1,3}$\footnote{miele@na.infn.it}}

\affiliation{
    $^1$Universit\`a ``Federico II'', Dipartimento di Scienze Fisiche, Napoli, Italy \& INFN Sezione di Napoli\\
    $^2$Department of Physics and Astronomy, University of Aarhus, Ny Munkegade, Bygn. 1520 8000 Aarhus Denmark \\
    $^3$Instituto de F\'{\i}sica Corpuscular (CSIC-Universitat de Val\`encia),
        Ed.\ Institutos de Investigaci\'on, Apartado de Correos 22085, E-46071 Val\`encia, Spain.
}

\date{\today}

\begin{abstract}
Annihilation of Dark Matter usually produces together with gamma
rays comparable amounts of electrons and positrons. The $e^+e^-$
gyrating in the galactic magnetic field then produce secondary
synchrotron radiation which thus provides an indirect mean to
constrain the DM signal itself. To this purpose, we calculate the
radio emission from the galactic halo as well as from its expected
substructures  and we then compare it with the measured diffuse
radio background. We employ a multi-frequency approach using data in
the relevant frequency range 100 MHz--100 GHz, as well as the WMAP
Haze data at 23 GHz. The derived constraints are of the order
$\sigva= 10^{-24}$\,cm$^3$s$^{-1}$ for a DM mass $m_{\chi}=100$\,GeV
sensibly depending however on the astrophysical uncertainties, in
particular on the assumption on the galactic magnetic field model.
The signal from single bright clumps is instead largely attenuated
by diffusion effects and offers only poor detection perspectives.

\end{abstract}
 \maketitle

\section{Introduction}
Cosmology and Astrophysics provide nowadays a compelling evidence of
the existence of Dark Matter (DM)
\cite{Komatsu:2008hk,Bertone:2004pz}. Nevertheless, its nature still
remains elusive, and Dark Matter constituents have escaped a direct
detection in laboratory so far. Promising candidates are DM
particles produced in thermal equilibrium in the early universe, the
so-called Weakly Interacting Massive Particles (WIMPs).
Theoretically, models of WIMPs naturally arise, for example, in SUSY
as the Lightest Super-symmetric Particle or as the Lightest
Kaluza-Klein Particle in the framework of extra-dimensions. These
candidates are self-conjugate and can thus annihilate in couples to
produce as final states: neutrinos, photons, electrons, light nuclei
(as wells as their antiparticles), etc., which can in principle be
detected.

Among the indirect DM detection channels, gamma-ray emission
represents one of the most promising opportunity due to the very low
attenuation in the interstellar medium, and to its high detection
efficiency. See for example Ref.s
\cite{Bertone:2004pz,Jungman:1995df,Bergstrom:2000pn} for a review
of this extensively studied issue. The expected neutrino detection
rates are generally low although  forthcoming km$^3$ detectors offer
some promising prospect \cite{Bergstrom:1998xh,Barger:2001ur}.
Finally, positrons and protons strongly interact with gas, radiation
and magnetic field in the galaxy and thus the expected signal
sensibly depends on the assumed propagation model
\cite{Baltz:1998xv,Hooper:2004bq,Donato:2003xg}. However, during the
process of thermalization in the galactic medium the high energy
$e^+$ and $e^-$ release secondary low energy radiation, in
particular in the radio and X-ray band, that, hence, can represent a
chance to look for DM annihilation. Furthermore, while the
astrophysical uncertainties affecting this signal are similar to the
case of direct $e^+$, $e^-$ detection, the sensitivities are quite
different, and, in particular, the radio band allows for the
discrimination of tiny signals even in a background many order of
magnitudes more intense.

Indirect detection of DM annihilation through secondary photons
has received recently an increasing attention, exploring the
expected signature in X-rays
\cite{Bergstrom:2006ny,Regis:2008ij,Jeltema:2008ax}, at radio
wavelengths
\cite{Blasi:2002ct,Aloisio:2004hy,Tasitsiomi:2003vw,Zhang:2008rs}
, or both
\cite{Colafrancesco:2005ji,Colafrancesco:2006he,Baltz:2004bb}. In
the following we will focus our analysis on the radio signal
expected from the Milky Way (MW) halo and its substructures. It is
worth noticing that the halo signal has been recently discussed in
Ref.s \cite{Hooper:2007kb,Hooper:2008zg,Grajek:2008jb} in
connection to the WMAP Haze, which has been interpreted as a
signal from DM annihilation. In this concern we will take in the
following a more conservative approach, by assuming that the
current radio observations are entirely astrophysical in origin,
and thus deriving constraints on the possible DM signal. The main
point will be the use of further radio observations besides the
WMAP ones, in the wide frequency range 100 MHz-100 GHz, and a
comparison of the achievable bounds. Furthermore, the model
dependence of these constraints on the assumed astrophysical
inputs will be analyzed. We will also discuss the detection
perspectives of the signal coming from the brightest DM
substructures in the forthcoming radio surveys.

The paper is organized as follows: in section \ref{AstrInput} we
will discuss the astrophysical inputs required to derive the DM
signal such as the structure of the magnetic field, the DM spatial
distribution and the radio data employed to derive the
constraints. In section \ref{synchsignal} we describe in detail
the processes producing the DM radio signal either when it is
originated from the halo or from the substructures. In section IV
we present and discuss our constraints, while in section V we
analyze the detection sensitivity to the signal coming from the
single DM clump. In section VI we give our conclusions and
remarks.

\section{Astrophysical Inputs}\label{AstrInput}

\subsection{Dark matter distribution}\label{DMD}

Our knowledge of the DM spatial distribution on galactic and
subgalactic scales has greatly improved thanks to recent high
resolution zoomed N-body simulations
\cite{Diemand:2005vz,Diemand:2006ik,Kuhlen:2008aw,Springel:2008cc}.
These simulations indicate that for the  radial profile of the
galactic halo the usual Navarro-Frank-White (NFW) distribution
\cite{Navarro:1996gj}
\begin{equation}
    \label{NFW}
    \rho(r)=\frac{\rho_{\text{h}}}{\frac{r}{r_{\text{h}}}\left(1+\frac{r}{r_{\text{h}}}\right)^2}
    \,\,\, ,
\end{equation}
still works as a good approximation over all the resolved scales. We
will thus use this profile in the following. Note, anyway, that this
choice is quite conservative with respect to other proposed
profiles like the Moore profile \cite{Moore:1999nt}, which exhibits
an internal cusp $\propto r^{-1.5}$ that would give in principle a
divergent DM annihilation signal from the center of the halo.
Observationally the situation is more uncertain: Baryons generally
dominate the gravitational potential in the inner kpc's and fitting
the data thus requires to model both the baryon and DM component at
the same time. The NFW profile is in fair agreement with the
observed Milky Way rotation curve \cite{Klypin:2001xu}, although,
depending on the employed model, it is possible to find an agreement
for many different DM profiles (see also \cite{Bertone:2004pz} and
references therein). We emphasize, however, that the various
profiles differ mainly in the halo center (for $r\alt 1$ kpc) where
the uncertainties, both in numerical simulations and from
astrophysical observations are maximal. Thus, our analysis which
explicitly excludes the galactic center, does not crucially  depend
on the choice of the profile.

A problem related to the profile of Eq.\ref{NFW} is that the mass
enclosed within the radius $r$ is logarithmically divergent. A
regularization procedure is thus required to define the halo mass.
Following the usual conventions we define the mass of the halo as
the mass contained within the \emph{virial radius} $r_{\rm{vir}}$,
defined as the radius within which the mean density of the halo is
$\delta_{\rm{vir}}=200$ times the mean \emph{critical cosmological
density} $\rho_{\text{cr}}$ which, for a standard cosmological model
($\Omega_{\text{m}}\simeq0.3, \Omega_\Lambda\simeq0.7$
\cite{Komatsu:2008hk}) is equal to $\rho_{\text{cr}}\simeq 5\times
10^{-6}$GeV\,$c^{-2}$\,cm$^{-3}$. The parameters describing the halo are then
determined imposing the DM density to be equal to $\rho_S=0.365$ GeV
$c^{-2}$ cm$^{-3}$ near the Solar System, at a galactocentric
distance of $R_{\text{S}}=8.5$ kpc.

Simulations, however, predict a DM distribution sum of a smooth halo
component, and of an additional clumpy one with total masses roughly
of the same order of magnitude. Hereafter we will assume for the
mass of the Milky Way
$M_{\text{MW}}=M_{\text{h}}+M_{\text{cl}}=2\times 10^{12}M_{\odot}$,
where  $M_{\text{h}}$ and $M_{\text{cl}}$ denote the total mass
contained in the host galactic halo and  in the substructures
(subhaloes) distribution, respectively. The relative normalization
is fixed by imposing that subhaloes in the range $\ms{7}$ ,
$\ms{10}$ have a total mass amounting to 10\% of $M_{\text{MW}}$
\cite{Diemand:2005vz}. Current numerical simulations can resolve
clumps with a minimum mass scale of $\sim\ms{6}$. However, for WIMP
particles, clumps down to a mass of $\ms{-6}$ are expected
\cite{Hofmann:2001bi,Green:2005fa}. We will thus consider a clump
mass range between $\ms{-6}$ and $\ms{10}$.

Finally, to fully characterize the subhalo population we will assume
a mass distribution $\propto m_{\text{cl}}^{-2}$ and that they are
spatially distributed following the NFW profile of the main halo.
The mass spectrum number density of subhaloes, in galactocentric
coordinates $\vec{r}$, is thus given by
\begin{equation}
\frac{{\d} n_{\text{cl}}}{{\d} m_{\text{cl}}}(m_{\text{cl}},\vec{r})= A
\left( \frac{m_{\text{cl}}}{M_{\text{cl}}} \right)^{-2} \left(
\frac{r}{r_{\text{h}}} \right)^{-1}\left(1+\frac{r}{r_{\text{h}}}\right)^{-2} \
,
\end{equation}
where $A$ is a dimensional normalization constant. The above
expression assumes some approximations: for example, a more
realistic clump distribution should take into account tidal
disruption of clumps near the galactic center. Numerical simulations
suggest also that the radial distribution could be somewhat
anti-biased with respect to the host halo profile. However, with our
conservative assumptions the DM annihilation signal is dominated by
the host halo emission within up to $20^\circ-30^\circ$ from the
galactic center so that the details of the sub-dominant signal from
the clumps have just a slight influence on the final results. Recent
results also show that mass distribution seems to converge to
$m_{\text{cl}}^{-1.9}$ rather than $m_{\text{cl}}^{-2}$
\cite{Kuhlen:2008aw}. This would also produce only a minor change in
the following results.

Following the previous assumptions the total mass in DM clumps of
mass between $m_1$ and $m_2$ results to be
\begin{eqnarray}\nonumber
M(m_1,m_2)&=&\int {\d}\vec{r}\, \int_{m_1}^{m_2} \,m_{\text{cl}}\,
\frac{{\d} n_{\text{cl}}}{{\d} m_{\text{cl}}}(m_{\text{cl}},\vec{r}) \, \, {\d}m_{\text{cl}} \\
\nonumber
          &=&4\pi\left[\ln{(1+c_{\text{h}})}-\frac{c_{\text{h}}}{1+c_{\text{h}}}\right]
             \left(A \, r_{\text{h}}^{3} \, M_{\text{cl}} \right)\\
          &\times& \ln{\left(\frac{m_2}{m_1}\right)}M_{\text{cl}}   \ ,
\end{eqnarray}
where $c_{\text{h}} \equiv r_{\rm vir}/r_{\text{h}}$ denotes the host halo
concentration; while their number is
\begin{eqnarray}\nonumber
N(m_1,m_2) &=&\int {\d}\vec{r}\, \int_{m_1}^{m_2} \, \frac{{\d}
n_{\text{cl}}}{{\d} m_{\text{cl}}}(m_{\text{cl}},\vec{r}) \, \,
{\d}m_{\text{cl}}
\\\nonumber
           &=&4\pi\left[\ln{(1+c_{\text{h}})}-\frac{c_{\text{h}}}{1+c_{\text{h}}}\right]
             \left(A \, r_{\text{h}}^{3} \, M_{\text{cl}} \right)\\
           & \times& \left(\frac{M_{\text{cl}}}{m_1}-\frac{M_{\text{cl}}}{m_2}\right) \ .
\end{eqnarray}
Imposing the normalization condition
$M(10^{7}M_{\odot},10^{10}M_{\odot})=10\% M_{\text{MW}}$, we
finally get for the mass due to the entire clumps distribution:
\begin{equation}
M_{\text{cl}}=M(10^{-6}M_{\odot},10^{10}M_{\odot})\sim 53.3\,\%
M_{\text{MW}}     \ ,
\end{equation}
while for the number of these clumps we obtain
\begin{equation}
N(10^{-6}M_{\odot},10^{10}M_{\odot}) \sim 2.90\times10^{17}\, .
\end{equation}
Finally by using the previous constraints one can fix the values
of free parameters $r_{\text{h}}$, $\rho_{\text{h}}$ and $A$, hence obtaining
$r_{\text{h}}=14.0$ kpc, that corresponds to a halo concentration of
$c_{\text{h}}=14.4$, $\rho_{\text{h}}=0.572$\,GeV $c^{-2}$ cm$^{-3}$ and $A=1.16
\times 10^{-19}$ kpc$^{-3}$ M$_{\odot}^{-1}$.

A further piece of information is required to derive the
annihilation signal from the clumps, namely how the DM is
distributed inside the clumps themselves. We will assume that each
clump follows a NFW profile as the main halo with $r_{\text{cl}}$
and $\rho _{\text{cl}}$ replacing the corresponding quantities of
Eq.\ref{NFW}. However, for a full characterization of a clump,
further information on its concentration $c_{\text{cl}}$ is
required. Unluckily, numerical simulations are not completely helpful
in this case, since we require information about the structure of
clumps with masses down to $\ms{-6}$, far below the current
numerical resolution. Analytical models are thus required. In the
current cosmological scenario \cite{Komatsu:2008hk} structures
formed hierarchically, {\it via} gravitational collapse, with
smaller ones forming first. Thus, naively,  since the smallest
clumps formed when the universe was denser, a reasonable expectation
is $c_{\text{cl}} \propto (1+z_f)$, where $z_f$ is the clump
formation redshift. Following the model of ref.
\cite{Bullock:1999he} we will thus assume $c_{\text{cl}}=c_1
\left(\frac{m_{\text{cl}}}{M_{\odot}}\right)^{-\alpha}$ with
$c_1=38.2$ and $\alpha=0.0392$. With this concentration the
integrated DM annihilation signal from all the substructures
dominates over the smooth halo component only at about $30^\circ$
from the galactic center (see section \ref{synchsignal}), so that
the constraints on the DM signal do not crucially depend on the
unresolved clumps signal, coming basically only from the smooth halo
component. However, given the large uncertainties in the models,
larger contributions from the unresolved population of clumps are in
principle possible considering a different parametrization of the
concentration (see for example the various models considered in
\cite{Pieri:2007ir, Kuhlen:2008aw}). We will not investigate further
this possibility here. An enhancement of the clumps signal is also
possible considering different choices of the clump profile other
than the NFW: Differently from the case of the halo, in fact, the
clump signal depends sensibly from the chosen profile and a Moore
profile or an Isothermal profile can in principle enhance the signal
of several orders of magnitude. Also in this case we choose to quote
conservative constraints and we will not consider these
possibilities further.

\subsection{Galactic Magnetic Field}
The MW magnetic field is still quite uncertain especially near the
galactic center. The overall structure is generally believed to
follow the spiral pattern of the galaxy itself with a
normalization of about $\sim 1\mu$G near the solar system.
Eventually, a toroidal or a dipole component is considered in some
model.

\begin{figure}
\vspace{1pc}
\begin{center}
\begin{tabular}{c}
\includegraphics[width=.6\columnwidth, angle=90]{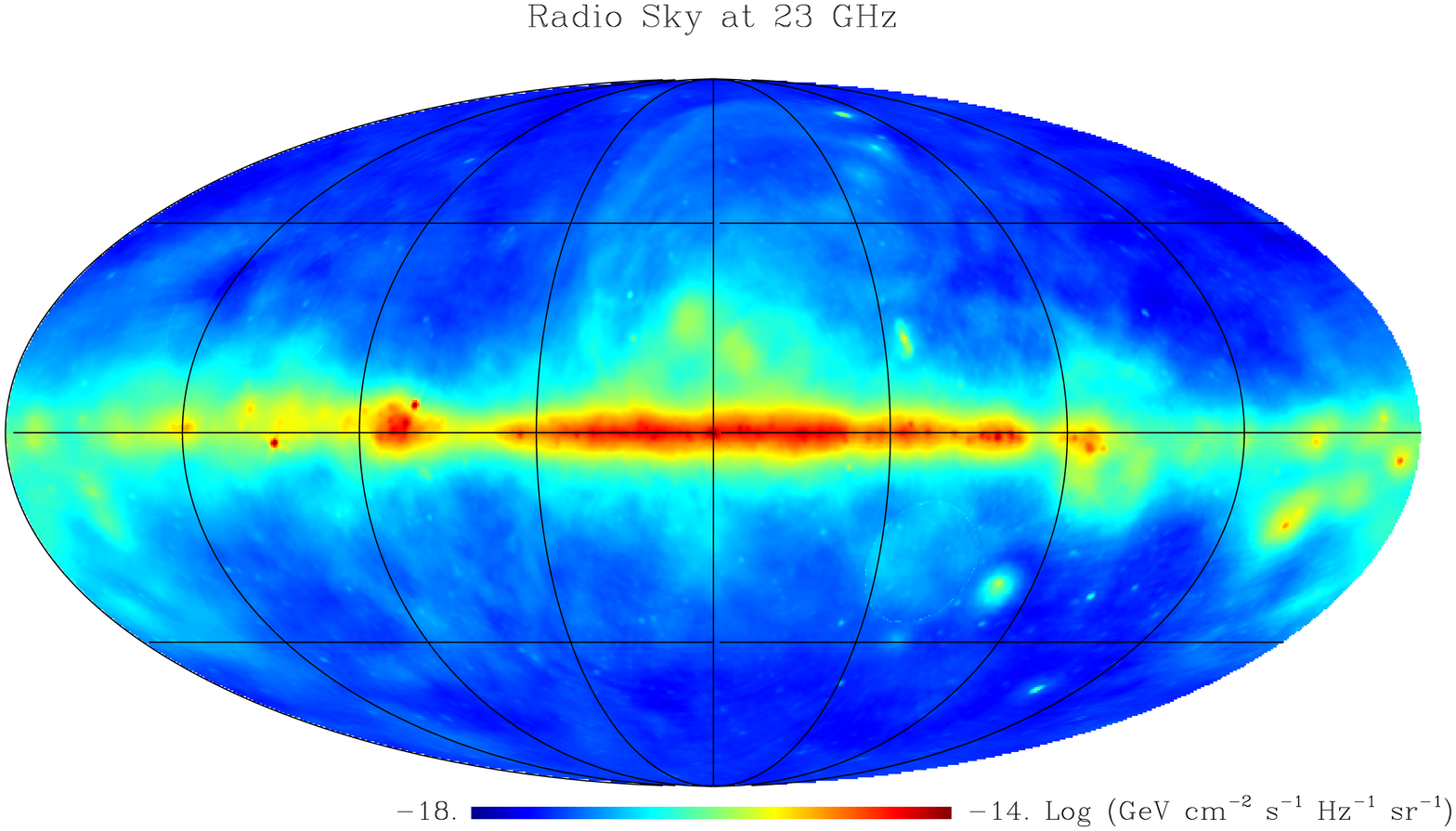}  \\
\includegraphics[width=.6\columnwidth,
    angle=90]{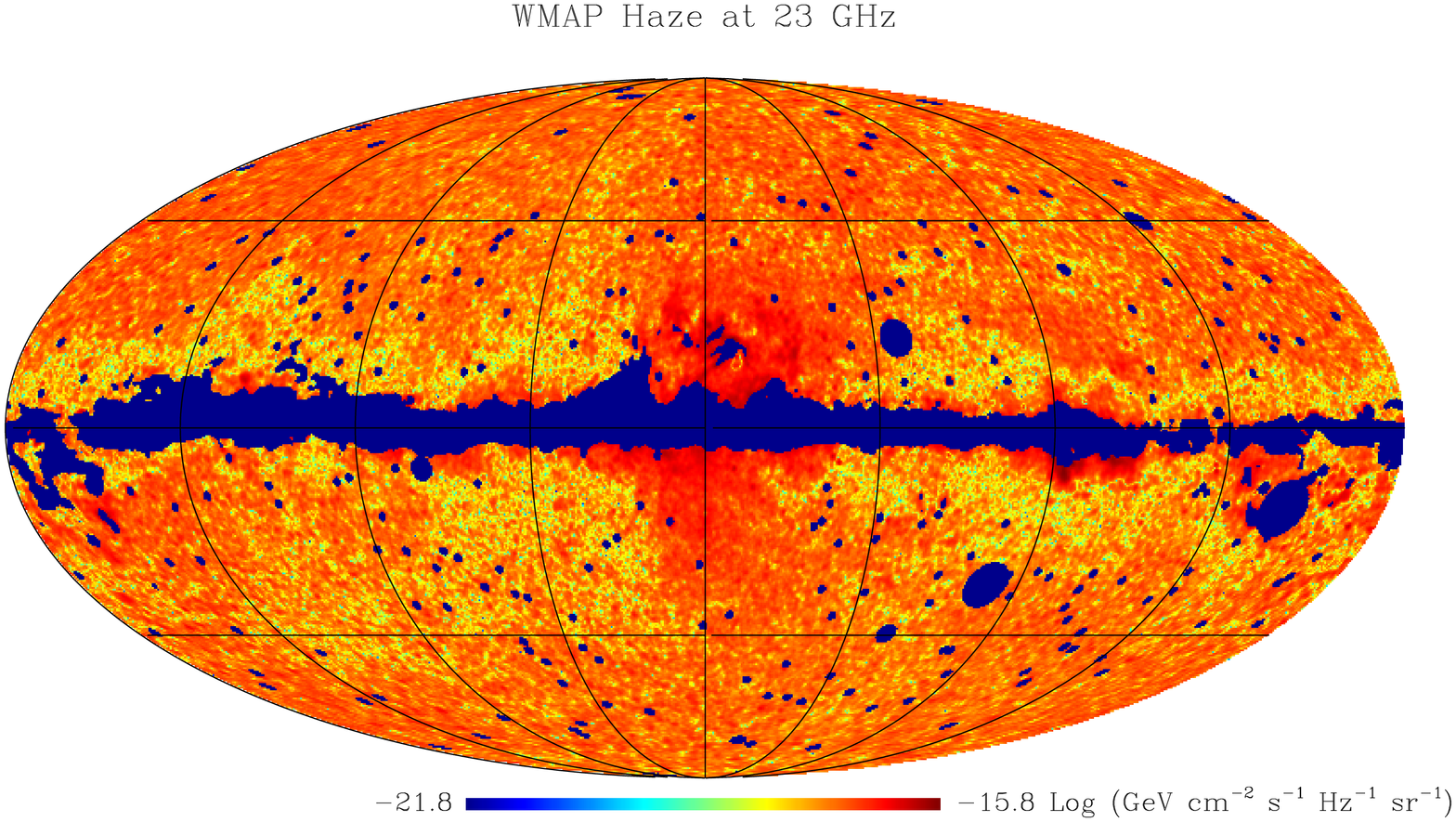} \\
\end{tabular}
\end{center}
        \caption{Sky map of galactic foregrounds at the frequency of
        23 GHz (top), and of the residual map showing the WMAP Haze (bottom).\label{fig:Skymaps}}
\end{figure}

We will consider in the following the Tinyakov and Tkachev model
(TT) \cite{Tinyakov:2001ir} which is a fair representative of the
available descriptions of MW magnetic field. Within this model , the
field shows the typical spiral pattern, an exponential decrease
along the $z$ axis and a $1/R$ behavior in the galactic plane. The
field intensity in the inner kiloparsecs is constant at about
$7\mu$G. We will use the slightly modified parametrization of this
model as described in \cite{Kachelriess:2005qm}. Higher
normalizations are in principle possible considering more complex
structures as for example a dipole or a toroidal component
\cite{Prouza:2003yf}. Indeed some recent analyses
\cite{Sun:2007mx,Noutsos:2008jw} including new available data seems
to favor the presence of these further structures. We will thus
consider as possible also an ``high normalization model'' that we
simply parameterize as a constant 10 $\mu$G field. This choice is
also motivated by a comparison with the results of
\cite{Hooper:2007kb,Hooper:2008zg} where the same magnetic field is
used.

Further, beside the regular component, the galactic magnetic field
presents a turbulent random component. The r.m.s. intensity of this
component is generally expected of the same order of magnitude of
the regular one, but both its spatial distribution and its spectrum
are poorly known, thus here we neglect its effects. Naively, this
random component is expected to affect the synchrotron maps that we
will show in the following  producing a blurring of the otherwise
regular pattern. Also, the random component  contributes to increase
the overall normalization of the field. Thus without this component
the synchrotron signal is slightly underestimated so that we can
regard this choice as conservative.

\subsection{Radio Data}
In the following we will derive  constraints on the DM emission
comparing the expected diffuse emission from the smooth halo and the
unresolved population of clumps with all sky observation in the
radio band. In the frequency range 100 MHz-100 GHz where the
DM synchrotron signal is expected, various astrophysical processes
contribute to the observed diffuse emission. Competing synchrotron
emission is given by Cosmic Ray electrons accelerated in supernovae
shocks dominating the radio sky up to $\sim$ 10 GHz. At higher
frequencies the Cosmic Microwave Background (CMB) and its
anisotropies represent the main signal. However, thanks to the very
sensitive multi-frequency survey by the WMAP satellite, this signal
(which represents thus a background for DM searches) can be modeled
in a detailed way and can thus be removed from the observed radio
galactic emission \cite{Tegmark:2003ve}. Other processes
contributing in the 10-100 GHz range are given by thermal
bremsstrahlung (free-free emission) of electrons on the galactic
ionized gas, and emission by small grains of vibrating or spinning
dust.

In the following our approach will be  to compare the DM signal with
the observed radio emission where only the CMB is modeled and
removed. For this purpose we use the code described in
\cite{deOliveiraCosta:2008pb} where most of the radio survey
observations in the range 10 MHz-100 GHz are collected and a scheme
to derive interpolated, CMB cleaned sky maps at any frequency in
this range is described.

A more aggressive approach would be of course to try to model and
subtract also the remaining emissions (synchrotron, free-free, dust)
in order to compare the expected DM signal with the residual radio
map. This is indeed the approach followed in
\cite{Finkbeiner:2003im,Dobler:2007wv} where residual maps at the 5
WMAP frequencies are derived using spatial templates for the various
expected  astrophysical components. The residual maps then exhibit
the feature called the WMAP Haze, which has been indeed interpreted
as radio emission related to DM annihilation
\cite{Finkbeiner:2004us,Hooper:2007kb}. However, the fit procedure
used for the Haze extraction is crucial, and using more degrees of
freedom to model the foregrounds as performed by the WMAP team
\cite{Gold:2008kp} fails in finding the feature. We will anyway show
in the following for comparison the constraints derived using the
Haze residual map at the WMAP frequency of 23 GHz \cite{HazeSkymap}.
A map of the Haze and of the 1 GHz emission is shown in
Fig.\ref{fig:Skymaps}. We will see however that within our
conservative approach comparable or better constraints can be
obtained thanks to the use of multi-frequency information. For a
given DM mass, in fact, 23 GHz is generally not the best frequency
to use and better constraints are instead obtained using
observations at lower frequencies even without further foreground
modeling.

Definitely, a detailed foreground modeling at all radio frequencies
would clearly give much stronger constraints on the DM signal and/or
eventually confirm the DM nature of the WMAP haze. To this purpose
consistent progress will be achieved in the next years with the new
high quality data coming from the PLANCK mission and from low
frequency arrays like LOFAR and SKA.

\section{DM Synchrotron Signal}\label{synchsignal}

\subsection{Particle Physics}
In a standard scenario where WIMPs experience a non exotic thermal
history, a typical mass range for these particles is $ 50 \,
\textrm{GeV} \alt m_\chi \alt 1 \, \textrm{TeV}$, while a simple
estimate for their (thermally averaged) annihilation cross section
yields $\sigva = 3 \times 10^{-27} \textrm{cm}^3\textrm{s}^{-1}/ \,
\Omega_{\text{cdm}} h^2$ \cite{Jungman:1995df}, giving $\sigva
\approx 3 \times 10^{-26} \textrm{cm}^3\textrm{s}^{-1}$ for
$\Omega_{\text{cdm}} h^2 \approx0.1$ as resulting from the latest
WMAP measurements \cite{Komatsu:2008hk}. However, this naive
relation can fail badly if, for example, coannihilations play a role
in the WIMP thermalization process \cite{Griest:1990kh}, and a much
wider range of cross sections should be considered viable. In this
work we consider values of $m_\chi$ from about 10 GeV to about 1 TeV, and $\sigva$
in the range ($10^{-26}$-$10^{-21}) \textrm{cm}^3\textrm{s}^{-1}$

The $e^+e^-$ annihilation spectrum for a given super-symmetric
WIMP candidate can be calculated for example with the DarkSUSY
package \cite{Gondolo:2004sc}. However, the final spectrum has
only a weak dependence on the exact annihilation process with the
channels $\chi\chi\rightarrow ZZ, W^+W^-, q\bar{q}$ giving
basically degenerate spectra. For leptonic channels like the
$\tau^+\tau^-$ decaying mode the spectrum differs significantly,
although this channel has generally a quite low branching ratio.
For simplicity we will assume hereafter full decay into $q\bar{q}$
channel, hence $e^-$ ($e^+$) will be emitted by decaying muons
(anti-muons) produced in pions decays. In this framework, the
resulting $e^+$, $e^-$ spectrum can be written as a convolution,
namely
\begin{eqnarray}\nonumber
\der{N_e}{E_e}(E_e)=\int_{E_e}^{m_\chi c^2}{\d}E_{\mu}\der{N_e^{(\mu)}}{E_e}(E_e,E_{\mu})\\
                    \times \int_{E_{\mu}}^{E_{\mu}/\xi}{\d}E_{\pi} W_{\pi}(E_{\pi})
                      \der{N_{\mu}^{(\pi)}}{E_\mu}(E_{\pi})
\end{eqnarray}
with $ \xi=\left(m_{\mu}/m_{\pi}\right)^2 \ , $ where
\begin{eqnarray}
\der{N_e^{(\mu)}}{E_e}(E_e,E_{\mu}) = \frac{2}{E_{\mu}}\left[
\frac{5}{6}-\frac{3}{2}\left(\frac{E_e}{E_{\mu}}\right)^2
+\frac{2}{3}\left(\frac{E_e}{E_{\mu}}\right)^3  \right],
\label{mudec}  \\
\der{N_{\mu}^{(\pi)}}{E_\mu}(E_{\pi}) =
\frac{1}{E_{\pi}}\frac{m_{\pi}^2}{m_{\pi}^2-m_{\mu}^2}\,\, ,
\,\,\,\,\,\,\,\,\,\,\,\,\,\,\,\,\,\,\,\,\,\,\,\,\,\,\,\,\,\,\,\,\,\,\,\,\,\,\,\,\,\,\,\,\,\,\,\,\,\,\,\,\,
\label{pidec} \\
W_{\pi}(E_{\pi}) = \frac{1}{m_\chi
c^2}\frac{15}{16}\left(\frac{m_\chi
c^2}{E_{\pi}}\right)^{\frac{3}{2}}\left(1-\frac{E_{\pi}}{m_\chi
c^2}\right)^2 . \label{had}
\end{eqnarray}
In particular, Eq. (\ref{mudec}) is the electron (positron)
spectrum produced in the muon (anti-muon) decay $\mu^-\rightarrow
e^-\nu_{\mu}\bar{\nu}_e$ ($\mu^+\rightarrow
e^+\bar{\nu}_{\mu}\nu_e$). Eq. (\ref{pidec}) stands for the
$\mu^-$ ($\mu^+$) spectrum from $\pi^-\rightarrow
\mu^-\bar{\nu}_{\mu}$ ($\pi^+\rightarrow \mu^+\nu_{\mu}$) decay
process, and, finally, Eq. (\ref{had}) provides a reasonable
analytical approximation of the spectrum of pions from $q\bar{q}$
hadronization \cite{Hill1983}. It is worth noticing that to be
more accurate Eq. (\ref{had}) should be substituted by a numerical
calculation, which however results not necessary for the aim of
the present paper as discussed in the following.

In this approximation the final electron (positron) spectrum can
be cast in a simple polynomial form of the ratio $E_e/\mchi$:
\begin{equation}
    \label{dNdE}
    \der{N_e}{E_e}(E_e)=\frac{1}{\mchi}\sum_{j\in
    J}^{}a_j\left(\frac{E_e}{\mchi}\right)^j \, ,
\end{equation}
where $J=\{-\frac{3}{2},-\frac{1}{2},0,\frac{1}{2},2,3\}$ and the
coefficients $a_j$ are listed in Table \ref{tab:ajvalues}.

\begin{table}[b]
    \centering
    \caption{$a_j$ values}
    \vspace{1mm}
\begin{tabular}{ccc}
\hline
\textbf{coefficient}  &  \textbf{analytical} & \textbf{numerical} \\
\hline
$a_ {-3/2}$  &  $\frac{65}{189}\frac{1-\xi^{3/2}}{1-\xi}$                                                               & $0.456$       \\
$a_ {-1/2}$  &  $-\frac{66}{7}\frac{1}{1+\xi^{1/2}}$                                                                    & $-5.37$       \\
$a_ {0}$     &  $\frac{25}{36}\frac{\xi^2-18\,\xi+8\,\xi^{1/2}+9}{(1-\xi)\xi^{1/2}}$                    & $10.9$        \\
$a_ {1/2}$   &  $9\frac{1-\xi^{-1/2}}{1-\xi}$                                                                                                   &   $-6.77$     \\
$a_ {2}$     &  $-\frac{3}{28}\frac{5\,\xi^2-42\,\xi+72\,\xi^{1/2}-35}{(1-\xi)\xi^{1/2}}$           &   $0.969$     \\
$a_ {3}$     &  $\frac{1}{189}\frac{35\,\xi^2-270\,\xi+424\,\xi^{1/2}-189}{(1-\xi)\xi^{1/2}}$   &   $-0.185$    \\
\hline
\end{tabular}
    \label{tab:ajvalues}
\end{table}

The main advantage of using the above analytical approximation
instead of a more accurate numerical input is that, as will be clear
in the next section, most of the observables for the radio emission
will be expressed in an analytical form as well. This, in turn, is
of help for a better understanding of the physical results.
Nevertheless, the difference with the complete numerical calculation
turns to be small, arising only for quite low
electron energies, and thus for very low radio frequencies. At low
energies, in fact, the analytical form has an asymptotic behavior
$\propto E_e^{-1.5}$ while the numerical spectrum has a turn down.
From a comparison with the numerical output from DarkSUSY for a 100
GeV WIMP with 100\% branching ratio into $b\bar{b}$ the analytical
form is a fair approximation until $E_e\approx1$ GeV, which for a magnetic field $B\sim10 \, \mu$G translates
into a minimum valid frequency $\nu=10$--$100$ MHz, thus below the
frequency window we are going to explore (see Eq.\ref{peaknu} below).

\subsection{Electrons equilibrium distribution}
Dark matter annihilation injects electrons in the galaxy at the
constant rate
\begin{equation} \label{inj1}
Q(E_e,r)=\frac{1}{2}\left(\frac{\rho(r)}{m_{\chi}}\right)^2 \sigva
\der{N_e}{E_e}\ .
\end{equation}
On the other hand, the injected electrons loose energy interacting
with the interstellar medium and diffuse away from the production
site. In the limit in which convection and reacceleration
phenomena can be neglected, the evolution of the $e^+e^-$ fluid is
described by the following diffusion-loss equation
\cite{Moskalenko:1997gh,Strong:1998pw,Strong:1998fr}
\begin{eqnarray}
\frac{\partial}{\partial t}\der{n_e}{E_e} &=&
\vec{\bigtriangledown} \cdot \bigg[K(E_{e},\vec{r})
\vec{\bigtriangledown} \frac{dn_{e}}{dE_{e}} \bigg] +
\frac{\partial}{\partial E_{e}}
\bigg[b(E_{e},\vec{r})\der{n_e}{E_e}  \bigg]\nonumber\\
&&+Q(E_{e},\vec{r}), \label{dif}
\end{eqnarray}
where ${\d}n_{e}/{\d}E_{e}$ stands for the number density of $e^+$, $e^-$
per unit energy, $K(E_{e},\vec{r})$ is the diffusion constant, and
$b(E_{e},\vec{r})$ represents the energy loss rate. The diffusion
length of electrons is generally of the order of a kpc (see section
\ref{singleclump}) thus for the diffuse signal generated all over
the galaxy, and thus over many kpc's, spatial diffusion can be
neglected. This is not the case for the signal coming from a single
clump for which the emitting region is much smaller than a kpc. We
will further analyze this point in section \ref{singleclump}. By
neglecting diffusion, the steady state solution can be expressed as
\begin{equation}\label{steady}
\der{n_e}{E_e}(E_{e},\vec{r}) = \frac{\tau}{E_{e}}
\int_{E_{e}}^{m_\chi c^2} \!\!\!\!\! dE_{e}' \,\,
Q(E_{e}',\vec{r})\, ,
\end{equation}
where $\tau=E_e/b(E_e,\vec{r})$ is the cooling time, resulting from
the sum of several energy loss processes that affect electrons. In
the following we will consider synchrotron emission and Inverse
Compton Scattering (ICS) off the background photons (CMB and
starlight) only, which are the faster processes and thus the ones
really driving the electrons equilibrium. Other processes, like
synchrotron self absorption, ICS off the synchrotron photons, $e^+
e^-$ annihilation, Coulomb scattering over the galactic gas and
bremsstrahlung are generally slower. They can become relevant for
extremely intense magnetic field, possibly present in the inner
parsecs of the galaxy \cite{Aloisio:2004hy}, and thus will be
neglected in this analysis.
\begin{figure}
\includegraphics[width=1.05\columnwidth, angle=0]{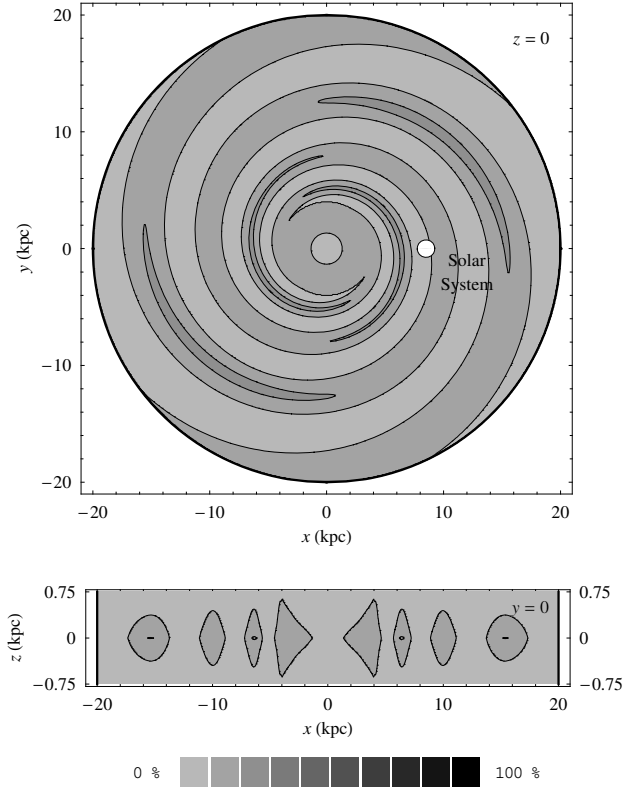}
\vspace{-0pc} \caption{Projections of the galaxy in the $xy$ and
$xz$ planes showing the fractional synchrotron contribution to the
$e^+e^-$ total energy losses for TT model \cite{Tinyakov:2001ir} of
GMF and Galprop model \cite{Porter:2005qx} of ISRF. The synchrotron
losses contribute up most to 20\,\% reaching its maximum at the
center of the magnetic arms. In the remaining regions, included the
galactic center, ICS is dominating. \label{fig:synchotr}}
\end{figure}

For synchrotron emission the energy loss is given by (for ex. see
\cite{Longair}) $b_{\text{syn}}(E_e) = 4/3 \, c\sigma_{\text{T}} \gamma^2
\beta^2 U_B$  with $U_B= B^2/2\mu_0$ the magnetic energy density
so that the time scale of the energy loss is:
\begin{equation}
\label{inj3} \tau_{\text{syn}}
          = \tau^0_{\text{syn}}{\B}^{-2}\left(\frac{E_e}{\textrm{GeV}}\right)^{-1}
\end{equation}
with $\tau^0_{\text{syn}} = 3.95\times 10^{17} \textrm{s}\,$.

Similarly, for Inverse Compton emission the energy loss is given by
$b_{\text{ICS}}(E_e) = 4/3 \, c\sigma_{\text{T}} \gamma^2 \beta^2 U_{\text{rad}}$.
The relevant radiation background for ICS is given by an
extragalactic uniform contribution consisting of the CMB with
$U_{\text{CMB}}=8\pi^5(kT)^4/15(hc)^3\approx 0.26$ eV/cm$^3$, the
optical/infrared extragalactic background and the analogous
spatially varying galactic contribution, the Interstellar Radiation
Field (ISRF). For the latter we use as template the Galprop
distribution model \cite{Porter:2005qx} which reduces to the
extragalactic one at high galactocentric distances. In this model,
the ISRF intensity near the solar position is about 5 eV/cm$^3$, and
reaches values as large as 50 eV/cm$^3$ in the inner kpc's. With
this model the ICS is always the the dominant energy loss process,
also near the galactic center (see Fig.\ref{fig:synchotr}). We thus
have
\begin{equation}
\tau_{\text{ICS}} =\tau^0_{\text{ICS}}
\left(\frac{U_{\text{rad}}(\vec{r})}{\textrm{eV}/\textrm{cm}^3}\right)^{-1}
\left(\frac{E_e}{\textrm{GeV}}\right)^{-1}\, ,
\end{equation}
with $\tau^0_{\text{ICS}}=9.82\times 10^{15} \textrm{s}$.

Finally, considering both the energy losses
$b_{\text{tot}}=b_{\text{syn}}+b_{\text{ICS}}$ we have
\begin{eqnarray}\label{tauloss}
\tau(E_e,\vec{r})&=&\Ee^{-1}\mu(\vec{r})\, \tau^0_{\text{syn}}\, ,   \\
\mu(\vec{r})&=&\left[\left(\frac{B(\vec{r})}{\mu
\textrm{G}}\right)^2+\frac{\tau^0_{\text{syn}}}{\tau^0_{\text{ICS}}}
\,
\frac{U_{\text{rad}}(\vec{r})}{\textrm{eV}/\textrm{cm}^3}\right]^{-1}\,
,
\end{eqnarray}
with the  function $\mu(\vec{r})$ enclosing the whole spatial
dependence.

\begin{figure}[!t]
 \centering
\includegraphics[width=.99\columnwidth]{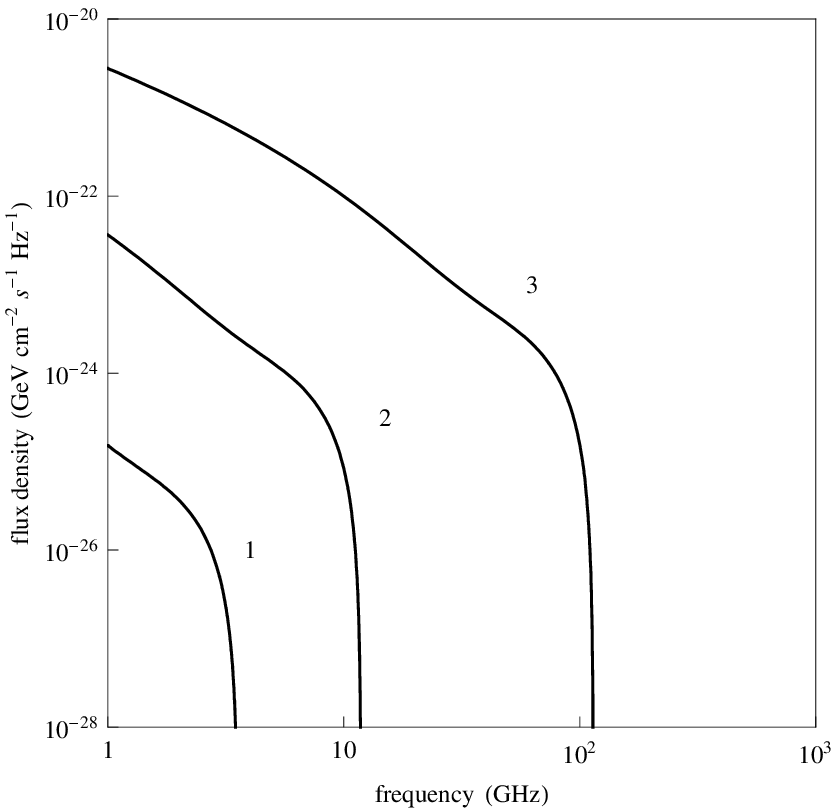}
\caption{Synchrotron flux density from three different clumps. The
neutralino parameters are $\mchi=100$ GeV and
$\sigva=3\times10^{-26}$ cm$^{3}\,$s$^{-1}$, while the relevant
clumps parameters are shown in table \ref{tab61}.\label{fdPlot}}
\end{figure}

By substituting the above expressions into Eq.(\ref{steady}) we
get the following equilibrium distribution for electrons
\begin{eqnarray}
\label{eqdistrfin} \der{n_e}{E_e}&=&\frac{\sigva
\tau^0_{\text{syn}}}{2\,m_{\chi} c^2}\,\mu(\vec{r})
                 \left(\frac{\rho(r)}{m_{\chi}}\right)^2\nonumber\\
              & &\times \sum_{k\in \mathcal{K}} b_k \mchiq^{-k}
                 \Ee^{k-1},
\end{eqnarray}
being $\mathcal{K}=J\cup\{-1\}$, $b_k=-a_k/(k+1)$, if $k\neq-1$,
while $b_{-1}=\sum_{j\in J}a_j/(j+1)$.

\begin{table}[b]
    \centering
\begin{tabular}{cccccc}\hline
clump& $d_{\text{cl}}$ & $r_{\text{cl}}$ & $\rho _{\text{cl}}$ &
$B$ & flux density at 1 GHz \\ \hline \#&
kpc&kpc&GeV$c^{-2}$cm$^{-3}$&$\mu$G&GeV\,cm$^{-2}$s$^{-1}$Hz$^{-1}$\\
\hline
1& 14.2 & 0.180 & 6.51 & 0.0962 & $1.70\times10^{-25}$ \\
2& 4.71 & 0.181 & 6.50 & 0.320 & $4.55\times10^{-23}$ \\
3& 5.50 & 0.188 & 6.404 & 3.08 & $2.66\times10^{-21}$ \\ \hline
\end{tabular}
\caption{Parameters of the example clumps chosen in Fig.
\ref{fdPlot}.\label{tab61}}
\end{table}

\subsection{Synchrotron spectrum}
The synchrotron spectrum of an electron gyrating in a magnetic field
has its prominent peak at the resonance frequency
\begin{equation}
    \label{peaknu}
\nu=\nu_0\left(\frac{B}{\mu
\textrm{G}}\right)\left(\frac{E_e}{\textrm{GeV}}\right)^2 \ ,
\end{equation}
with $\nu_0=3.7\times10^6\,\textrm{Hz}$. This implies that, in
practice, a $\delta$--approximation around the peaks works
extremely well. Using this \emph{frequency peak} approximation,
the \textit{synchrotron emissivity} can be defined as
\begin{equation}
    \label{synemiss}
j_{\nu}(\nu,\vec{r})
                =\der{n_e}{E_e}(E_e(\nu),\vec{r})\,\der{E_e(\nu)}{\nu}\,b_{\text{syn}}(E_e(\nu),\vec{r}).
\end{equation}
This quantity is then integrated along the line of sight for the
various cases to get the final synchrotron flux across the sky.

\begin{figure}[!t]
    \vspace{1pc}
    \centering
    \includegraphics[width=.99\columnwidth]{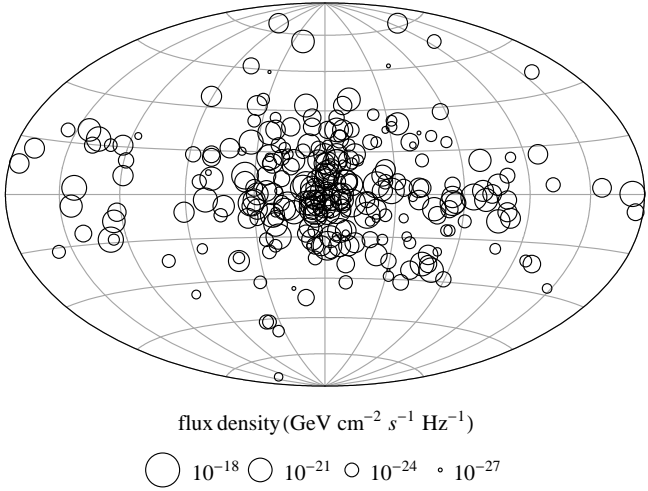}
    \caption{Sky map at the frequency of
    1 GHz for a realization of clumps distribution. For each clump, the circle radius is proportional to the logarithm
    of radio flux.\label{fig:SyncPlot}}
\end{figure}

\subsubsection{Single clump signal}
According to the description of previous section \ref{DMD}, let us
consider a clump of mass $m_{\text{cl}}$, whose center of mass is
placed at $\vec{R}_{\text{cl}}$ and with a sufficiently small
size. In this case it is possible to neglect the spatial variation
of $\mu(\vec{r})$ inside the clump itself, and thus the flux
$I_{\nu}$ can be calculated as:
\begin{equation}
 \label{flux}
I_{\nu}(\nu, \vec{R}_{\text{cl}})=\frac{1}{4 \pi \,
d_{\text{cl}}^2}\int
{\d}\vec{r}\,j_{\nu}(\nu,\vec{R}_{\text{cl}}+\vec{r})\,
 \, ,
\end{equation}
with $d_{\text{cl}}$ the distance between the observer and the
clump. This can be rewritten as
\begin{eqnarray}
\label{Inunu} I_{\nu}(\nu, \vec{R}_{\text{cl}}) =
I_{\nu}^0\,\mu(\vec{R}_{\text{cl}})\sum_k A_k
{\left(\frac{B(\vec{R}_{\text{cl}})}
{\mu\textrm{G}}\right)}^{1-\frac{k}{2}}\left(\frac{\nu}{\textrm{Hz}}\right)^{\frac{k}{2}}  \\
A_k(m_{\chi}) = b_k \mchiq^{-k} \nuz^{-\frac{k}{2}-1},\phantom{\hspace{.2\columnwidth}}
\end{eqnarray}
with
\begin{eqnarray}
\label{Inu0}
\frac{I_{\nu}^0}{\textrm{GeV cm$^{-2}$s$^{-1}$ Hz$^{-1}$}}=\hspace{.4\columnwidth}\nonumber\\
=2.57\times10^{-12} \left(\frac{\mchi}{100\,\textrm{GeV}}\right)^{-3} \frac{\sigva}{10^{-26}\,\textrm{cm$^3$s$^{-1}$}}\nonumber\\
\times \rclo^3 \dcl^{-2} \rhocl^2 \, .
\end{eqnarray}
\vspace{0.5pc}

\begin{figure}[!t]
 \centering
 \includegraphics[width=0.85\columnwidth]{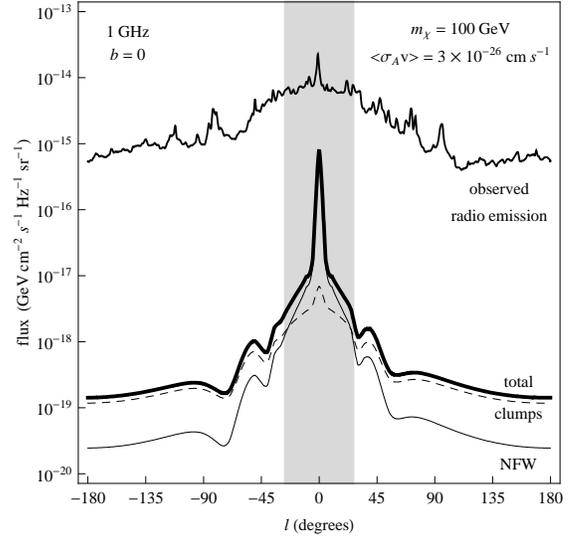}
 \caption{DM synchrotron profile for the halo and unresolved
 substructures and their sum at 1 GHz for $m_\chi=100$ GeV  and
 $\sigva=3\times10^{-26}$\,cm$^3$s$^{-1}$. The astrophysical observed
 emission at the same frequency is also shown. The gray band indicates
 the angular region within which the DM signal from the host halo dominates
 over the signal from substructures modeled as in section \ref{DMD}.}
 \label{fig:HVSCL}
\end{figure}

Fig. \ref{fdPlot} shows some examples of signal, produced by three
clumps of our simulation. An important feature to notice is that the
synchrotron signal sensibly depends on the magnetic field both in
the normalization and in the covered frequency range. In particular,
the signal frequency cutoff, remnant of the energy spectrum cutoff
near $m_\chi$, depends on $B$ following Eq.\ref{peaknu}.

\vspace{0.5pc}

Fig. \ref{fig:SyncPlot} shows instead the positions and radio
intensities for a realization of the clumps distribution with masses
$m_{\text{cl}}>\moMS{7}$. It can be seen that all the clumps with a
non negligible signal lie near the galactic plane where most of the
galactic magnetic field is concentrated. Few clumps are visible at
high latitude just because of projection effects, being located very
near and slightly up or below the solar position with respect to the
galactic plane.


\subsubsection{Diffuse signals}


The diffuse halo signal is similarly given by the integral along the
line of sight of Eq.(\ref{synemiss})
\begin{equation}
\frac{\d^2 I_{\nu}}{{\d}l\,{\d}b}=\frac{\cos{b}}{4\pi}
\int_0^{\infty} j_{\nu} \, {\d}s \, ,
\end{equation}
where $(l,b)$ are coordinates on the sphere and $s$ the line of
sight coordinate. To calculate the total contribution from the
substructures, instead,  we have to sum over all haloes
\begin{equation}
\frac{{\d}^2I_{\nu}^{\text{unr}}}{{\d}l\,{\d}b}
        =\cos{b}
\int {\d}m_{\text{cl}} \int {\d}s\, s^2 \, \frac{{\d}
n_{\text{cl}}}{{\d}
m_{\text{cl}}}(m_{\text{cl}},\vec{r})\,I_{\nu}^{\text{res}}(\nu,
\vec{r})\, ,
\end{equation}
with $I_{\nu}^{\text{res}}$ given by Eq.(\ref{Inunu}) and
$\vec{r}=\vec{r}(s,l,b)$.

Interestingly, the sum of the two diffuse contributions can be
rewritten as
\begin{equation}
\frac{{\d}^2
I_{\nu}^{\text{DM}}}{{\d}l\,{\d}b}=\frac{\cos{b}}{4\pi}\int
j_{\nu}^{\text{DM}}\, {\d}s \, ,
\end{equation}
where
\begin{widetext}
\vspace{-0pc}
\begin{eqnarray}\nonumber
j_{\nu}^{\text{DM}}&=& \frac{1}{4} \mchiq^{-3}
\frac{\sigva}{\textrm{cm$^3$s}^{-1}}
\left\{\left[\frac{\rho_{\text{h}}/\textrm{GeV$c^{-2}$cm$^{-3}$}}{(r/r_{\text{h}})(1+r/r_{\text{h}})^2}\right]^2+
\frac{\rho_{\text{CL}}/\textrm{GeV$c^{-2}$cm$^{-3}$}}{(r/r_{\text{h}})(1+r/r_{\text{h}})^2}\right\}\\
&&\times\,\mu(\vec{r})\sum_k A_k(m_{\chi}){\left(\frac{B(\vec{r})}{\mu\textrm{G}}\right)}^{1-k/2}\nuq^{k/2}
\textrm{GeV\,cm$^{-3}$s$^{-1}$Hz$^{-1}$sr$^{-1}$}\, .
\end{eqnarray}
\end{widetext}
Thus, from the point of view of DM annihilation the unresolved
clumps signal behaves like a further smooth NFW component with the
same scale radius of the halo profile, but with a different
\emph{effective density}
$\rho_{\text{CL}}=0.604\,\textrm{GeV$c^{-2}$cm}^{-3}$, and with an
emissivity simply proportional to the density profile instead of its
square.

We see that the halo component dominates in the central
region of the galaxy, where
\begin{equation}
\frac{r}{r_{\text{h}}}\left(1+\frac{r}{r_{\text{h}}}\right)^2<\frac{\rho_{\text{h}}^2}{\rho_{\text{CL}}}
\quad\Rightarrow\quad r<4.39\,\textrm{kpc}
\end{equation}
which corresponds to a disk of radius 27.3 degrees (see fig. \ref{fig:HVSCL}).

\section{DM Annihilation constraints}
\begin{figure}[!t]
    \includegraphics[width=0.60\columnwidth, angle=90]{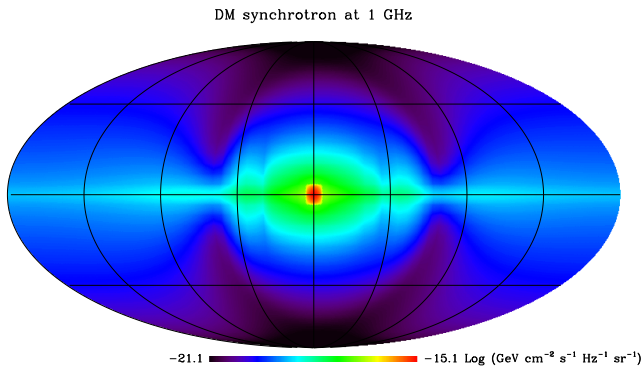}
    \caption{Sky map of the galactic radio signal generated by the DM smooth
    halo and  unresolved clumps
    at the frequency of 1\,GHz for $m_\chi=100$ GeV and
$\sigva=3\times10^{-26}$cm$^3$s$^{-1}$. The peculiar shape of the
signal, pinched approximately at $\pm 30^\circ$ and $\pm 60^\circ$,
reflects basically the structure of the magnetic field as seen in
projection from the Solar System, where the observer is located
(compare also with Fig.\ref{fig:synchotr}). The galactic center
region and the first few magnetic arms are visible as regions of
high magnetic field intensity and hence prominent synchrotron
emission.}
  \label{fig:halocontours}
\end{figure}

The pattern and intensity of the DM radio map resulting from the sum
of the contributions from the smooth halo and unresolved clumps is
shown in Fig. \ref{fig:halocontours} for $m_\chi=100$ GeV and
$\sigva=3\times10^{-26}$cm$^3$s$^{-1}$. Similar maps are obtained at
different frequencies and different $m_\chi$ and $\sigva$ to obtain
DM exclusion plots. For our analysis we use a small mask covering a
$15^\circ$$\times$$15^\circ$ region around the galactic center where
 energy loss processes other than synchrotron and ICS start
possibly to be relevant. We include the galactic plane although this
region  has basically no influence for the constraints on the DM
signal.

\begin{figure*}[!t]
\begin{center}
\begin{tabular}{cc}
\includegraphics[width=.85\columnwidth]{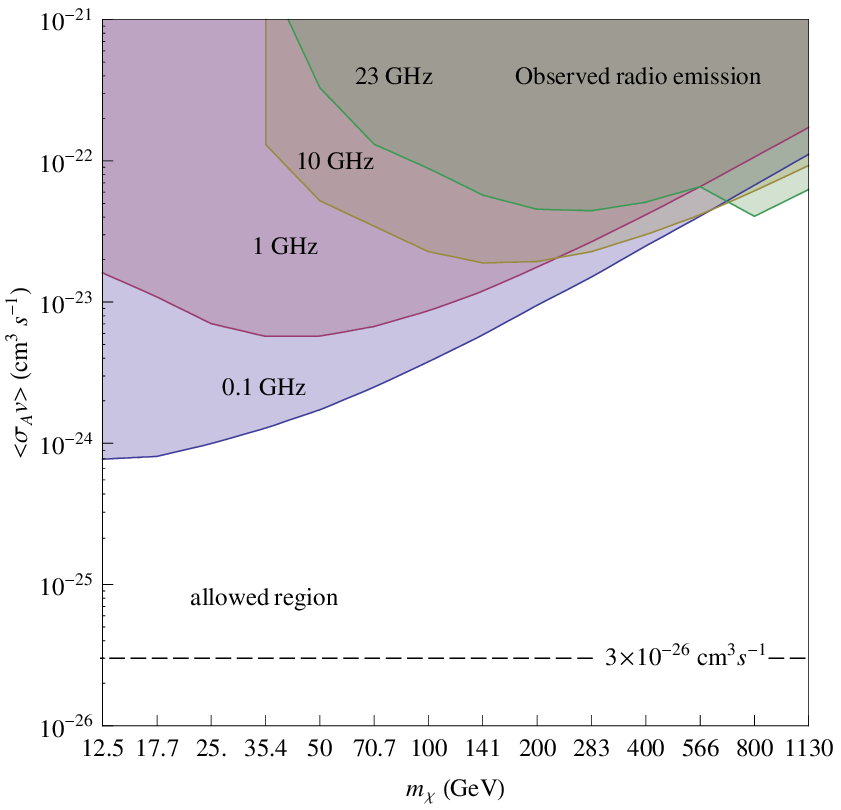} &
\hspace{3pc}
\includegraphics[width=.85\columnwidth]{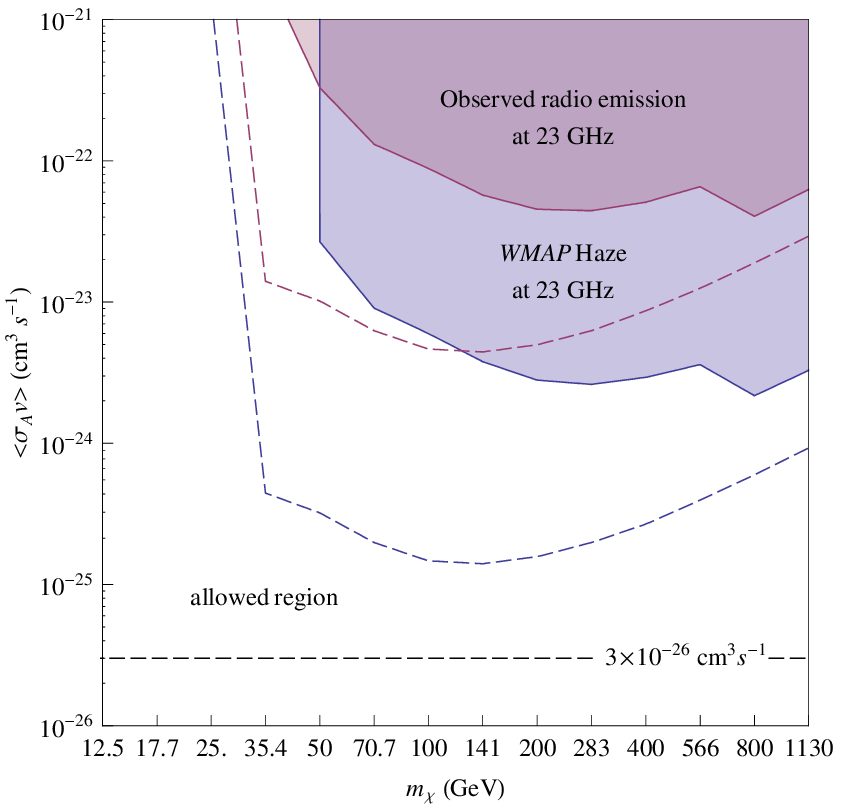}\\
\end{tabular}
\end{center}
\caption{(Left) Constraints in the  $m_\chi$-$\sigva$ plane for
various frequencies without assuming synchrotron foreground removal.
(Right) Constraints from the WMAP 23 GHz foreground map and 23 GHz
foreground--cleaned residual map (the WMAP Haze) for the TT model of
magnetic field (filled regions) and for a uniform 10 $\mu$G field
(dashed lines). } \label{fig:DMconstraints}
\end{figure*}

In Fig.\ref{fig:DMconstraints} we show the radio constraints on the
DM annihilation signal in the $m_\chi$--$\sigva$ plane for various
frequencies and various choices of the foreground. Several comments
are in order. First, we can see that, as expected, the use of the
haze at 23 GHz gives about one order of magnitude better constraints
with respect to the synchrotron foregrounds at the same frequency.
However, using also the information at other frequencies almost the
same constraints can be achieved. This information in particular is
complementary giving better constraints at lower DM masses. This is
easily understood since a smaller DM mass increases the annihilation
signal ($\propto m_\chi^{-2}$) at smaller energies, and thus smaller
synchrotron frequencies. In particular, the constraints improve of
about one order of magnitude at $m_\chi\sim100$ GeV from 23 GHz to 1
GHz while only a modest improvement is achieved considering further
lower frequencies as 0.1 GHz. This saturation of the constraints is
due to the frequency dependence of the DM signal, that below 1 GHz
becomes flatter than the astrophysical backgrounds so that the
fraction of contribution from DM is maximal at about 1 GHz. Further,
the constraints show a threshold behavior given basically by
Eq.\ref{peaknu} which settles a maximum emitted radio frequency for
a given DM mass $m_\chi$. This threshold behavior is, for example,
clearly seen at 23 GHz in the right panel of
Fig.\ref{fig:DMconstraints} where only for masses above $\sim 40$
GeV the cross section is constrained.

Notice that although the astrophysical background  which we compare
with at 1 GHz is an interpolation, the derived constraints are still
valid given the smooth behavior and the broad frequency extent of
the DM signal, which does not exhibit narrow peaks at particular
frequencies. However, effective measurements have been performed for
example at 408 MHz and 1.4 GHz (see \cite{deOliveiraCosta:2008pb}).
Quoting our constraints at these exact frequencies would change the
results only slightly.

The DM signal has thus a broad frequency extent and also below 1 GHz
is still relevant. This is a potential problem for the DM
interpretation of the WMAP Haze given that, in the Haze extraction
procedure, the observed radio emission at 408 MHz is used as
template of the synchrotron background. In fact, naively, a DM
signal at 23 GHz should be relevant at 408 MHz as well, unless
either the DM mass or the magnetic field is so high to
shift the DM contribution to higher frequencies and making it
negligible at 408 MHz.

The second relevant point to notice is that the constraint depends
quite sensibly on the magnetic field assumptions. The constraints we
obtain with the TT model are generally almost two orders of magnitude
weaker with respect to the results reported in \cite{Hooper:2008zg}.
They are instead  more in agreement with \cite{Grajek:2008jb} where
Galprop has been employed to calculate the DM synchrotron signal.
For a closer comparison with \cite{Hooper:2008zg} we choose, as they
do, a constant magnetic field of 10\,$\mu$G although still
keeping the Galprop ISRF model. Even in this case our derived
constraints are a factor of 5 weaker (despite the inclusion of the
contribution from substructures). The remaining factor of 5 can be
finally recovered using a constant ISRF with
$U_{\text{rad}}=5$\,eV/cm$^3$ as assumed in \cite{Hooper:2008zg}. In
this case, in fact,  the smaller values of $U_{\text{rad}}$ reduces
the ICS losses enhancing in turn the synchrotron signal. It should
be said however that, while the magnetic field normalization is
still quite uncertain, the ISRF is instead more constrained and a
large variation with respect to the Galprop model seems unlikely.

The constraints shown in Fig.\ref{fig:DMconstraints} extend down to
\mbox{$\sim 10$ GeV}, which is somewhat the mass limit for a
conservative analysis. It is clear that for low masses the
constraints come more and more from lower frequencies: For example
for a WIMP of 30 GeV the data at 100 MHz are 2 orders of magnitude
more constraining than the data at 10 GHz. However, extremely low
frequencies are not experimentally accessible. For a WIMP of 1 GeV,
from Eq.\ref{peaknu} with a magnetic field of $\mathcal{O}$($\mu$G)
only frequencies $\alt 10$ MHz would be useful to place constraints
on the DM signal. Although observations at this frequency exist
\cite{deOliveiraCosta:2008pb}, in general the survey sky coverage is
quite incomplete and the data quality is non-optimal. Observations
in this very low frequency range should substantially improve with
the next generation radio arrays LOFAR and SKA. WIMP masses below 1
GeV still would produce observable synchrotron radiation at the
galactic center where the magnetic field is likely much higher than
$\mu$G scale (possibly $\mathcal{O}$(mG) ). This kind of analysis
would be however quite model dependent and would face further
background uncertainties.

\section{Single clumps detectability} \label{singleclump}
To have a reliable estimate of the sensitivity to a single clump
detection diffusion effects cannot be neglected. Although the
integrated synchrotron clump signal is given by Eq.(\ref{Inunu}),
the clumps appear extended rather than pointlike with a dimension
typically of several degrees. As a reasonable approximation we can
assume that the signal is spread over an area of radius equal to the
diffusion length of the electrons
$l_D=\sqrt{K(E_e)\tau_{\text{loss}}(E_e)}$, where $K$ is the
diffusion coefficient and $\tau_{\text{loss}}$ is the energy loss
time given by Eq.(\ref{tauloss}). We use for $K$ the Galprop model
\cite{Moskalenko:1997gh}
\begin{equation}
    K=K_0 \left( \frac{E_e}{E_{e0}} \right)^\delta \, ,
\end{equation}
with a reference energy $E_{e0}=3$ GeV, a Kolmogorov spectrum
$\delta=1/3$ and $K_0=10^{28}$cm$^2$/s.

Taking as reference the parameters of a very bright clump like the
\#3 in table \ref{tab61}, we get ($1\textrm{pc}=3\times10^{18}\textrm{cm}$)
\begin{equation}
    l_D = \sqrt{K(E_e)\Ee^{-1}\mu(\vec{x})\, \tau^0_{\text{syn}}}
    \approx 1 \, \textrm{kpc} \, ,
\end{equation}
for $E_e\approx 10$ GeV and for a radiation density
$U_{\text{rad}}\approx5$\,eV/cm$^3$. The energy losses are basically
dominated by ICS thus the result is almost independent of the
magnetic field value. Moreover the dependence on the electron
energy and the radiation density itself is very weak. Of course the
clumps will have a certain profile peaked in the center and will not
be perfectly smoothed all over $l_D$. However the dilution of the
signal in the much larger volume with respect to the region of
emission makes it quite hard to detect the clump. We can consider for
example the signal from a very bright clump at a distance of 5 kpc
with a flux of $10^{-22}$GeV\,cm$^{-2}$s$^{-1}$Hz$^{-1}$ at 20 GHz,
corresponding approximately to the characteristics of clump \#3 in
Fig.\ref{fdPlot}. With a dilution over 1 kpc, the clump emission is
seen under a steradian $A=\pi\alpha^2\approx0.1$ sr with
$\alpha\approx 10^\circ$, giving a diffuse clump flux of
$10^{-21}$GeV\,cm$^{-2}$s$^{-1}$Hz$^{-1}$sr$^{-1}$. The WMAP
sensitivity of about $10\,\mu$K translates into a flux sensitivity\footnote{At radio frequencies the Rayleigh-Jeans law
$F_\nu=2\nu^2/c^2 k_{\text{B}}T$ is employed to translate fluxes
into brightness temperatures} of $\sim
10^{-18}$GeV\,cm$^{-2}$s$^{-1}$Hz$^{-1}$sr$^{-1}$, meaning that the
expected, optimistic signal is about 3 order of magnitude below the
reach of the current sensitivity. The situation is only slightly
better at 150 MHz where the expected LOFAR sensitivity is 50 mK
\cite{Jelic:2008jg} i.e. $\sim 2\times
10^{-19}$GeV\,cm$^{-2}$s$^{-1}$Hz$^{-1}$sr$^{-1}$.

The chance of clump radio detection seems thus quite poor even
with the next generation experiments. On the other side, the fact
that the signal is anyway extended and not pointlike makes the
clump signal not really complementary to the diffuse component
sharing the same systematics with a much fainter signal. It is
likely thus that the a role for DM investigations in the radio
will be played basically by the diffuse signal.

\section{Summary and conclusions}
Using conservative assumptions for the DM distribution in our galaxy
we  derive the expected secondary radiation due to synchrotron
emission from high energy electrons produced in DM annihilation. The
signal from single bright clumps offers only poor sensitivities
because of diffusion effects which spread the electrons over large
areas diluting the radio signal. The diffuse signal from the halo
and the unresolved clumps is instead  relevant and can be compared
to the radio astrophysical background to derive constraints on the
DM mass and annihilation cross section.

Constraints in the radio band, in particular, are complementary to
similar (less stringent but less model dependent) constraints in the
X-ray/gamma band \cite{Mack:2008wu,Kachelriess:2007aj} and from
neutrinos \cite{Yuksel:2007ac}. Radio data, in particular, are more
sensitive in the GeV-TeV region while neutrinos provide more
stringent bounds for very high DM masses ($\agt 10$ TeV). Gammas,
instead, are more constraining for $m_\chi \alt 1$ GeV. The
combination of the various observations provides thus interesting
constraints over a wide range of masses  pushing the allowed window
significantly near the thermal relic possibility.

More into details, we obtain conservative constraints at the level
of $\sigva \sim 10^{-23}$\,cm$^3$s$^{-1}$ for a DM mass
$m_{\chi}=100$\,GeV from the WMAP Haze at 23 GHz. However, depending
on the astrophysical uncertainties, in particular on the assumption
on the galactic magnetic field model, constraints as strong as
$\sigva \sim 10^{-25}$\,cm$^3$s$^{-1}$ can be achieved.
Complementary to other works which employ the WMAP Haze at 23 GHz,
we also use  the information in a wide frequency band in the range
100 MHz-100 GHz. Adding this information the constraints become of
the order of $\sigva \sim 10^{-24}$\,cm$^3$s$^{-1}$ for a DM mass
$m_{\chi}=100$\,GeV. The multi-frequency approach thus gives
comparable constraints with respect to the WMAP Haze only, or
generally better for $m_\chi\alt100$ GeV where the best sensitivity
is achieved at $\sim$ GHz frequencies.

The derived constraints are quite conservative because no attempt to
model the astrophysical background is made differently from the case
of the WMAP Haze. Indeed, the Haze residual map itself should be
interpreted with some caution, given that the significance of the
feature is at the moment still debated and complementary analyses
from different groups (as the WMAP one) miss in finding a clear
evidence of the feature. In this respect, the multifrequency
approach will be definitely necessary to clarify the nature of
controversial DM signals as in the case of the WMAP Haze. Progresses
are expected with the forthcoming data at high frequencies from
Planck and at low frequencies from LOFAR and, in a more distant
future, from SKA. These surveys will help in disentangling the
various astrophysical contributions thus assessing the real
significance of the Haze feature. Further, the low frequency data in
particular, will help to improve our knowledge of the galactic
magnetic field. Progresses in these fields will provide a major
improvement for the interpretation of the DM-radio connection.

\vspace{1pc}
\section*{Acknowledgments}
We wish to thank P.D. Serpico for valuable comments and T. Di
Girolamo for a careful reading of the draft. Use of the publicly
available HEALPix software \cite{Gorski:2004by} is acknowledged.
G.M. acknowledges supports by the Spanish MICINN (grants
SAB2006-0171 and FPA2005-01269) and by INFN--I.S.Fa51 and PRIN 2006
``Fisica Astroparticellare: Neutrini ed Universo Primordiale" of
Italian MIUR.


\end{document}